\documentclass{article}

\usepackage[preprint]{neurips_2026}

\usepackage[T1]{fontenc}
\usepackage[utf8]{inputenc}
\usepackage{hyperref}
\hypersetup{
  pdftitle={The Alpha Illusion: Reported Alpha from LLM Trading Agents Should Not Be Treated as Deployment Evidence},
  pdfauthor={Yuxuan Ye, Jun Han, Ao Hu, Juncheng Bu, Yiyi Chen, Liangjian Wen, Danilo Mandic, Danny Dongning Sun, Xu Yinghui, Zenglin Xu}
}
\usepackage{url}
\usepackage{booktabs}
\usepackage{amsfonts}
\usepackage{amsmath}
\usepackage{amssymb}
\usepackage{nicefrac}
\usepackage{microtype}
\usepackage{xcolor}
\usepackage{array}
\usepackage{xspace}
\usepackage{graphicx}

\setcitestyle{numbers,square,comma,sort&compress}

\newcommand{\PPL}{\textsc{Parametric Prior Lock-in}\xspace}

\title{The Alpha Illusion: Reported Alpha from LLM Trading Agents Should Not Be Treated as Deployment Evidence}

\author{%
  Yuxuan Ye$^{1,*}$ \quad
  Jun Han$^{2,*}$ \quad
  Ao Hu$^{3}$ \quad
  Juncheng Bu$^{4}$ \quad
  Yiyi Chen$^{1}$ \\[2pt]
  Liangjian Wen$^{3}$ \quad
  Danilo Mandic$^{5,\dagger}$ \quad
  Danny Dongning Sun$^{6,\dagger}$ \quad
  Xu Yinghui$^{1,\dagger}$ \quad
  Zenglin Xu$^{1,\dagger}$ \\[4pt]
  $^{1}$Fudan University \quad
  $^{2}$Shanghai University of Finance and Economics \\
  $^{3}$Southwest University of Finance and Economics \quad
  $^{4}$Northeastern University \\
  $^{5}$Imperial College London \quad
  $^{6}$Peng Cheng Laboratory \\[4pt]
  \texttt{zenglinxu@fudan.edu.cn} \\[2pt]
  {\small $^{*}$Equal contribution. \quad $^{\dagger}$Corresponding authors.}
}

\begin{document}
\maketitle
\raggedbottom

\begin{abstract}
End-to-end LLM trading agents have moved quickly from research curiosity to a small ecosystem of named systems, including FinCon, FinMem, TradingAgents, FinAgent, QuantAgent, and FLAG-Trader. Several of these report headline Sharpe ratios that would be material if read at face value on a deployment desk, and associated benchmarks such as FinBen report trading-task Sharpe statistics in the same range. The gap between architecture research and deployment claim has been crossed too freely on both sides of the academia--industry divide. We take a position on that gap: \textbf{reported alpha from end-to-end LLM trading agents should not be treated as deployment evidence}. Before such returns can support claims of deployable trading capability, they must survive structural validity tests for temporal integrity, real-world frictions, counterfactual robustness, predictive calibration, numerical execution, and multi-agent disaggregation. Current public evidence cannot yet distinguish robust predictive ability from temporal contamination, unmodeled frictions, short-window Sharpe uncertainty, narrative fitting, and parametric priors. The problem is not only evaluative but structural. Language confidence is not tradable probability, narrative reasoning is not numerical execution, and model priors may become undisclosed implicit factor exposures. We contribute a minimum reporting protocol suite, P1--P6, with tiered applicability by claim strength, and a conservative modular alternative that uses LLMs as auditable information interfaces upstream of independent calibration, risk, and execution modules. Code and reproduction harness: \url{https://github.com/hj1650782738/Trading}.
\end{abstract}

\section{Introduction}
\label{sec:intro}

End-to-end LLM trading agents sit at the intersection of two communities whose evaluative languages have drifted apart. The quantitative-trading industry asks a deployment question: can a language-model agent generate net returns under live execution? Academia, over the past two years, has moved in a different but equally valid direction---exploring new architectures and expanding the design space. NeurIPS 2024 accepted FinCon~\citep{yu2024fincon} (manager--analyst hierarchy with portfolio decisions) on the main track; alongside FinMem~\citep{yu2023finmem}, TradingAgents~\citep{xiao2024tradingagents}, FinAgent~\citep{zhang2024finagent}, QuantAgent~\citep{xiong2025quantagent}, and FLAG-Trader~\citep{xiong2025flagtrader}, this body of work is best read as exploratory architecture research, which is the appropriate stance for a fast-moving area. The benchmarking side has moved in parallel: FinBen~\citep{xie2024finben}, a holistic financial-LLM benchmark on the NeurIPS 2024 Datasets and Benchmarks Track ($36$ datasets, $24$ tasks), reports a FinTrade subtask Sharpe that we use later as an evaluation anchor rather than as a trading framework. The benchmark-to-deployment gap is itself a known ML pathology: \citet{lipton2018troubling} catalogue how loose evaluation, narrative inflation, and metric--mechanism conflation recur across subfields, and trading is where that conflation is most concrete.

The difficulty arises one step downstream. Because no shared boundary separates ``exploratory architecture'' from ``deployable system,'' readers sometimes carry headline numbers from short, in-cutoff backtests with single-component cost models forward as deployment evidence. When industry attempts deployment and discovers the gap, the reaction overshoots into ``academic LLM-trading work cannot be relied on,'' and the two communities stop reading each other on the right terms: the careful exploratory contribution gets dismissed, and industry treats academia as out of touch. The conflation has concrete shape---a plausible investment narrative is not a tradable probability; a short positive return is not robust alpha; and a multi-agent debate pipeline is not aggregation of independent experts. Without protocols ruling out training-corpus contamination, unaccounted frictions, short-sample significance, miscalibration, numerical-execution errors, and parametric-prior exposure, backtestable but non-deployable agents enter the literature under ``high-Sharpe'' headlines, and teams arrive at real markets with miscalibrated expectations.

\textbf{We argue: until they pass structural validity tests, the alpha reported by end-to-end LLM trading systems should not be interpreted as evidence of deployable trading capability.} This position does not deny LLM value in finance; the strongest same-side evidence~\citep{lopezlira2023chatgpt} shows that LLMs can extract semantic signals from news headlines that correlate with short-term market reactions, which is precisely the modular use of LLMs we endorse. When evidence is insufficient to support a deployment interpretation, the safer design lets the LLM enter the trading pipeline through independent calibration, risk, execution, and audit modules rather than directly bearing final trading decisions.

Our aim is to defend the academic exploratory role rather than to police it. We make three contributions: (i) we unify temporal contamination, friction, multiple-testing, miscalibration, numerical execution, and parametric-prior issues into a single ``reported alpha is not deployment evidence'' framework; (ii) we propose a minimum reporting protocol P1--P6 tiered by claim strength, giving authors, reviewers, and deployment teams a common yardstick; (iii) we describe a conservative modular practice---LLMs as auditable information interfaces upstream of independent calibration, risk, and execution modules. The focus is evidentiary, not a new strategy or benchmark.

\begin{figure}[t!]
  \centering
  \includegraphics[width=\linewidth]{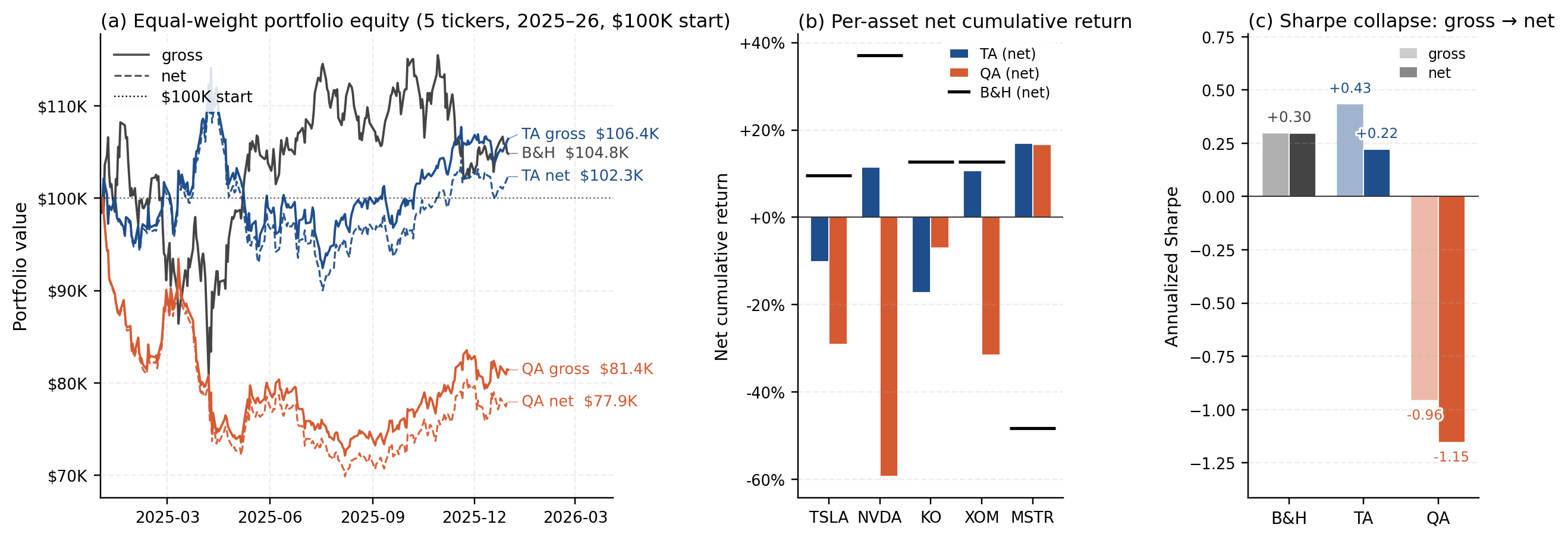}
  \caption{One-year reproduction of TradingAgents (TA) and QuantAgent (QA) vs.\ Buy-and-Hold (B\&H) (equal-weight TSLA/NVDA/KO/XOM/MSTR, \$100K start). ``Net'' charges commission, token cost, spread, and market impact; ``Gross'' is the pre-friction trajectory. Setup and per-asset breakdown in Appendix B.}
  \label{fig:portfolio}
\end{figure}

\section{Reported Alpha Does Not Yet Constitute Deployment Evidence}
\label{sec:eval-confounds}

Existing results from end-to-end LLM trading systems should not be dismissed: the returns they report \textbf{should first be understood as historical backtest or prototype evidence, not deployment evidence}. A short-window positive trajectory does not automatically imply transferable, auditable, deployable trading capability; reported alpha must first rule out three confounders: temporal contamination, real-world friction, and researcher degrees of freedom.

\subsection{What Positive Agent Results Do and Do Not Show}

These systems explore distinct engineering forms---layered memory, analyst/trader/risk role hierarchies, multimodal ingestion, manager--analyst CVaR risk control, HFT-style indicator/pattern/trend agents, and PPO fine-tuning with a Sharpe-increment reward---and make ``can an LLM trade directly'' a meaningful question. But the published evaluations sit on short, prompt-specific windows, with publication-time headline numbers that are uniformly strong (per-system windows and per-ticker Sharpe summary in Appendix A, Figure~\ref{fig:system-ticker}; FinCon~\citep{yu2024fincon} reaches per-ticker Sharpe $2.37$ and portfolio Sharpe $3.27$; QuantAgent reports $50.7$--$63.7\%$ directional accuracy on HFT bars). These numbers evidence ``a positive-return trajectory under the given protocol,'' not deployment-grade alpha (post-cutoff, point-in-time, dynamic-universe, full-friction). In the contamination-free real-market evaluation of \citet{chen2025stockbench}, most LLM agents struggle to consistently outperform passive buy-and-hold once temporal leakage and execution realism are controlled.

\subsection{Temporal Contamination: When Backtests Measure Memory Rather than Prediction}
\label{sec:temporal}

Temporal contamination is where LLM trading evaluation is harder than traditional quantitative backtesting. Traditional look-ahead bias usually comes from explicit data errors---using future prices, future earnings revisions, or post-hoc index constituents. But LLM contamination can be more subtle: even when the experimental code feeds only contemporaneously visible information, the model weights, post-training data, or retrieval corpora may already have absorbed post-event news, market post-mortems, price-path summaries, and year-end reviews. The model need not explicitly access future prices to leak future information through semantic memory.

The most direct quantitative anchor is~\citet{li2025profitmirage}: comparing LLM trading agents inside vs.\ outside their pretraining cutoff, FinMem's total return drops by $\approx 71.85\%$ and QuantAgent's Sharpe by $\approx 51.48\%$ once the window crosses the cutoff, directly quantifying the contamination component of reported alpha. \citet{merchant2025lookahead} mechanistically corroborate this with Divergence Decoding, which strips future semantic memory at inference time---supporting the more fundamental claim that \textbf{contamination is not only a prompt or retrieval-timestamp problem, but may have already entered model parameters and semantic associations}.

The contamination problem also cuts in the opposite direction. \citet{shah2025beyond} show that even within the reported cutoff, LLM financial knowledge is sharply biased by firm size and recency---LLMs answer revenue questions accurately for $54.17\%$ of large-cap firms in 2017 but only $6.32\%$ in 1995, with hallucination rates rising for larger and more recent firms; \citet{kang2023deficiency} document the same factual hallucination pattern in finance more broadly. So a strong backtest in a recent, mega-cap-heavy window is doubly fragile: it may both leak future information through semantic memory and rest on knowledge gaps for the smaller or older firms real deployment must also handle. Both directions point to the same conclusion: \textbf{LLM financial knowledge is not point-in-time reliable}.

We therefore argue: post-cutoff and contamination-free evaluation is not a bonus but a minimum prerequisite. If an end-to-end LLM trading system performs well only inside or under unclear contamination relative to the model's knowledge cutoff, the more plausible interpretation is ``the backtest performance may include historical semantic memory'' rather than ``the model learned to predict the market.''

\subsection{Real-World Friction: Better Prediction Does Not Imply Higher Net Return}
\label{sec:friction}

Trading systems are ultimately evaluated by net returns---not explanation quality, directional accuracy, or gross-return curves. Any gross-to-net translation obeys a minimum decomposition:
\begin{equation}
  \mathrm{PnL}_{\text{net}} \;=\; \mathrm{PnL}_{\text{gross}} \;-\; \sum_t \Big( c \cdot |\Delta w_t| \;+\; s_t \cdot |\Delta w_t| \;+\; \kappa \cdot |\Delta w_t|^{\beta} \Big),
  \label{eq:pnl_decomp}
\end{equation}
where $c$ is commission, $s_t$ is the bid--ask spread at time $t$, and $\kappa, \beta$ characterize temporary market impact ($\beta \approx 1$ under the standard \citet{almgren2000optimal} setup), with order-book resilience and execution-cost dynamics modeled by \citet{obizhaeva2013optimal}. The decomposition is descriptive, not optional: every term must be estimated, deducted, and audited, otherwise the gap between reported gross PnL and deployable net PnL remains open. For multi-agent LLM systems, the right-hand side must additionally include token costs and inference latency; returns are eroded not only by market friction but also by the system's own inference cost.

\citet{jang2025rl} provide a direct counterexample: after RL training, BTC market-state classification accuracy improves markedly, yet cumulative returns under simulated trading decline---the core implication of equation~\eqref{eq:pnl_decomp}: optimizing a gross-side proxy (classification accuracy, sharper direction, more coherent rationales, longer debates) does not optimize the net-side reality of equation~\eqref{eq:pnl_decomp}.

Across FinMem, TradingAgents, FinAgent, FinCon, and QuantAgent, $35$ of $40$ system\,$\times$\,friction-component cells are unmodeled (coverage matrix in Appendix B, Figure~\ref{fig:friction}); reported gross Sharpe is therefore an upper bound on any deployable net Sharpe. Figure~\ref{fig:portfolio} anchors this gap on a one-year, five-ticker reproduction of TradingAgents and QuantAgent: charging commission, token cost, spread, and market impact (full setup in Appendix B), portfolio Sharpe drops from $0.43\!\to\!0.22$ for TradingAgents and from $-0.96\!\to\!-1.15$ for QuantAgent, and both agents end below buy-and-hold.

In multi-agent debate scenarios the cost-eats-return problem becomes more systemic: \citet{zhang2025multiagent} report that across $36$ configurations ($4$ models $\times$ $9$ benchmarks), multi-agent debate wins under $20\%$ of the time, and adding rounds or agents does not improve---and may degrade---performance.

\subsection{Short Windows and Search Space: High Sharpe Does Not Stand Automatically}
\label{sec:short-window}

The third confound comes from short-window estimation and researcher degrees of freedom. Many LLM trading papers report high Sharpe ratios, high cumulative returns, or significant outperformance versus a benchmark over a few-month or single-year window. But the financial econometrics literature has long established that short-window Sharpe estimates carry substantial uncertainty, and repeated search over candidate signals, rule sets, and window choices inflates the false-positive rate.

\begin{figure}[t]
  \centering
  \includegraphics[width=0.95\linewidth]{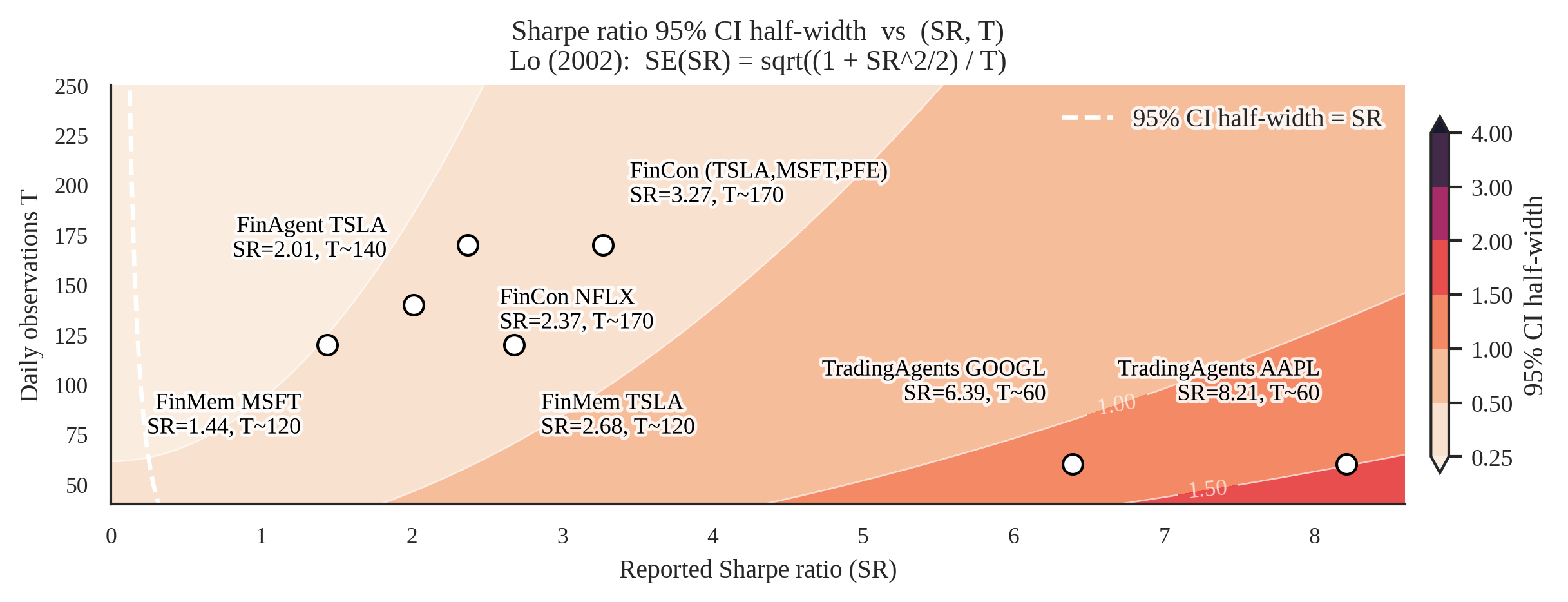}
  \caption{Heatmap of $95\%$ CI half-width of period Sharpe under \citet{lo2002sharpe} (eq.~\ref{eq:lo_se}, $K{=}1$), over $\widehat{SR} \in [0, 8]$ and $T \in [20, 500]$ daily observations. The dashed line marks where the $95\%$ CI just covers zero. Overlaid headline Sharpes from TradingAgents, FinMem, FinAgent, and FinCon~\citep{yu2024fincon} are illustrative (FinCon: single-stock NFLX $2.37$ and best portfolio $(\text{TSLA},\text{MSFT},\text{PFE})$ $3.27$); QuantAgent is omitted because it reports HFT directional accuracy rather than daily-frequency Sharpe. Annualized headlines use eq.~\ref{eq:lo_se} with $K{=}252$. The overlays are illustrative of reporting uncertainty rather than re-estimations of the original papers' confidence intervals, since original Sharpe conventions are not always comparable.}
  \label{fig:sharpe-ci}
\end{figure}

\citet{lo2002sharpe} gives the Sharpe standard-error approximation
\begin{equation}
  \mathrm{SE}(\widehat{SR}) \;\approx\; \sqrt{\tfrac{K + \widehat{SR}^{\,2}/2}{T}},
  \label{eq:lo_se}
\end{equation}
with $K = 1$ for period Sharpe and $K = 252$ for annualized-from-daily Sharpe; unless papers disclose which convention they use, headline Sharpes are not comparable across systems. Uncertainty is largest precisely in the headline-friendly regime of small $T$ and large $\widehat{SR}$. Figure~\ref{fig:sharpe-ci} plots the resulting $95\%$ CI half-width and overlays representative LLM-trading headlines; the FinBen GPT-4 FinTrade Sharpe of $1.51 \pm 1.08$ (std exceeding half the mean)~\citep{xie2024finben} sits in the same quadrant, so standard errors routinely exceed typical between-system gaps.

Researcher degrees of freedom worsen the problem. Traditional quant strategies have degrees of freedom in factor definitions, look-back windows, rebalance frequencies, transaction-cost assumptions, and portfolio weights; LLM systems add model versions, temperature, system prompts, few-shot examples, RAG corpora, memory length, agent personas, debate rounds, output-parsing rules, and tool-call strategies. The empirical asset pricing literature already shows that even with disciplined ML pipelines, monthly stock-level $R^2$ rarely exceeds $\approx 0.4\%$ and portfolio-level $R^2$ $\approx 1$--$2\%$~\citep{gu2020empirical}; against this ceiling, multi-decade factor-zoo work~\citep{harvey2016cross} long ago argued for a $t$-statistic hurdle near $3.0$ to control for multiple testing. \citet{novymarx2025llmharking} demonstrate that LLMs can be used to filter and produce full papers from over $30{,}000$ candidate signals, warning of industrialized HARKing.

\section{From Evaluation Confounds to Structural Mismatches}
\label{sec:prior-lockin}

A natural objection to Section 2 is that stricter post-cutoff data, full friction modeling, and statistical testing remedy everything. Better evaluation discipline is necessary but not sufficient: end-to-end LLM trading still faces three structural mismatches---language confidence is not tradable probability; financial narrative ability is not numerical execution; parametric priors in model weights can become undisclosed implicit factor exposures.

\subsection{Language Confidence Is Not Tradable Probability}
\label{sec:output-mismatch}

Trading systems require not a ``plausible-sounding'' explanation but a probability object that can be calibrated, sized, and risk-constrained. Position sizing, stop-loss thresholds, risk budgeting, and portfolio optimization all depend on some form of conditional probability or conditional return distribution. By contrast, an LLM's natural output comes from a next-token distribution over the language space, optimized to produce text that is more likely, more coherent, and more preference-aligned under the corpus distribution. Confidence in the language space and conditional probability in the financial-return space are not the same object.

Calibration is the quantifiable form of this distinction. \citet{kadavath2022knowing} show that LLMs' self-reported confidence about their own answers, while non-trivially informative, is systematically miscalibrated and degrades sharply on out-of-distribution inputs---a pattern directly inherited by any system that pipes verbal confidence into trade sizing. \citet{guo2017calibration} provides the widely used Expected Calibration Error:
\begin{equation}
  \mathrm{ECE} \;=\; \sum_{m=1}^{M} \frac{|B_m|}{n}\,\big|\mathrm{acc}(B_m) - \mathrm{conf}(B_m)\big|,
  \label{eq:ece}
\end{equation}
where $\{B_m\}_{m=1}^{M}$ partitions samples by predicted confidence. ECE measures the weighted average distance between ``claimed probability'' and ``empirical frequency''; any system that uses an LLM's self-reported confidence as a position input implicitly commits to $\mathrm{ECE} \approx 0$, and that commitment must be measured independently rather than substantiated by tone. \citet{guo2017calibration}'s findings come from discriminative image classifiers, so the cross-domain extrapolation here is indirect (consistent direction, but evidence drawn from a different domain); the financial domain has same-side empirics, most directly~\citet{lee2025knowledge}'s counterfactual interventions---even when reverse evidence is explicitly presented, self-reported model confidence and direction-flip behavior do not strictly cohere.

Our claim is not that \citet{guo2017calibration} directly proves LLM trading miscalibration, but that any system converting verbal confidence into position size inherits a measurable calibration obligation. \citet{shen2025calibrated} formalize the downstream side: calibration error in predictions fed to decision algorithms translates into bounded suboptimality of the induced action, making calibration a precondition for any decision-rule built on the predicted distribution.

\subsection{Financial Narrative Ability Is Not Numerical Execution}

The second mismatch is the gap between narrative reasoning and numerical execution. LLMs can explain complex financial concepts---options payoff structure, macro-event impact, balance-sheet changes, sector-rotation logic, or the intuition of a trading strategy. But trading systems ultimately execute numerical operations: computing risk exposure, estimating volatility, handling cash flows, controlling leverage, satisfying portfolio constraints, setting stops, and slicing orders.

TraderBench operationalizes this gap: it places financial agents in static-knowledge tasks, analytical-reasoning tasks, options trading, and adversarial crypto-trading scenarios, and evaluates them with realizable metrics including Sharpe, returns, drawdown, P\&L, Greeks, and risk management~\citep{yuan2026traderbench}. Its public summary shows that extended thinking substantially improves retrieval-style tasks but barely helps trading performance; in options tasks the models exhibit a clear conceptual-vs-computational gap---they can recognize strategies or explain concepts but struggle to quantify P\&L, Greeks, and risk exposure precisely. \citet{ma2025arena} provides a complementary same-side anchor: in a virtual zero-sum stock market with bid--ask interactions, LLM agents struggle with text-only numerical reasoning---they perform notably better when the same numerical state is rendered as a chart than when it is presented as text, indicating that the bottleneck is numerical execution rather than financial vocabulary. \textbf{This is precisely our point: speaking finance fluently is not the same as computing it correctly under real trading constraints.}

FinToolBench~\citep{lu2026fintoolbench} further shows that financial tool use must be evaluated on finance-critical dimensions such as call-level compliance, timeliness, intent alignment, and domain alignment, rather than binary execution success alone. We therefore treat tool-calling as a necessary safety patch, not a sufficient architectural fix.

\subsection{Parametric Priors as Undisclosed Implicit Factor Exposures}

The third mismatch occurs at the weight layer. An LLM is not a neutral processor of financial information. Its pretraining corpus contains historical news, market commentary, investment narratives, company coverage, social-media discussion, and ex-post post-mortems; preference optimization further rewards clear, helpful, explanatory, human-aligned answers. Together, these processes can produce stable default preferences when the model is faced with financial questions---for example, a tilt toward mega-cap tech, narratively interesting companies, or mainstream sectors.

We propose this as an explanatory framework rather than an established unified mechanism, called \textbf{\PPL}. It refers to: prior to any real-time input, the model weights already contain stable industry, size, style, or narrative tilts; these tilts shape the final trade recommendation but, unlike traditional factor exposures, are typically not explicitly disclosed, measured, or risk-constrained. A traditional quant portfolio with technology, mega-cap, momentum, quality, value, or contrarian exposure can be measured by factor regression, sector neutralization, and a risk model; in end-to-end LLM trading systems, style preference may hide inside natural-language reasoning and surface as ``apparently autonomous analysis.''

PPL admits a minimal counterfactual diagnostic. Following~\citet{lee2025knowledge}, we define the view-flip rate under reverse-evidence strength $\rho \in [0, 1]$ as
\begin{equation}
  \phi_{\text{vol}}(\rho) \;=\; \Pr\!\left[\,\hat{a}_{t+1} \neq \hat{a}_t \,\middle|\, e_t,\,\rho\,\right],
  \label{eq:flip}
\end{equation}
where $\hat{a}_t$ is the baseline view and $\hat{a}_{t+1}$ is the view updated with reverse evidence $e_t$ at strength $\rho$. We do not claim a Bayesian-rational agent must drive $\phi_{\text{vol}}\to 1$: strong reverse evidence may rationally call for reduced confidence or smaller position size rather than a sign flip. We claim instead a weaker monotonicity: a well-updating agent should exhibit a monotone response in at least one of (i) direction-flip rate, (ii) self-reported confidence, or (iii) realized position size as $\rho$ increases; an agent locked by parametric prior fails this monotonicity on all three axes simultaneously. \citet{lee2025knowledge}'s counterfactual interventions provide same-side evidence; Appendix D expands the supporting findings on sector tilts, cross-model homogeneity, and the ineffectiveness of persona prompts and multi-agent debate at removing the shared prior.

For trading systems, such tilts are not ordinary text bias but \emph{undisclosed implicit factor exposures}.

\subsection{Why Common Patches Do Not Solve the Problem in Isolation}
\label{sec:patches}

The four standard optimistic responses---scale, longer context / RAG, tool-calling, and multi-agent debate---are addressed individually in Section 6, Concerns 2--5; none alone supports an end-to-end deployment claim, and the benchmark-to-deployment gap is not unique to finance: web-agent and general LLM-agent benchmarks display the same overestimation under longer, friction-laden, adversarial trajectories~\citep{liu2024agentbench,zhou2024webarena}, with trading being the same pathology under harsher friction.

\section{Minimum-Evidence Protocols for Deployment Claims}
\label{sec:protocols}

\textbf{We argue: reported alpha from end-to-end LLM trading systems should not be treated as evidence of deployable trading capability until it survives structural validity tests.} When a paper interprets trading returns as deployment-ready alpha or autonomous trading ability, it must carry a heavier evidentiary burden. We propose six reporting protocols P1--P6 that we treat as a \emph{minimum necessary screening protocol}, not a sufficient certification: passing P1--P6 does not by itself imply real-world deployability---operational risk, model governance, order routing, capacity, tail-risk behavior, and adversarial market impact remain open. Rather, P1--P6 are a compact set of failure modes such that \emph{failing any one of them} is, on present evidence, sufficient to disqualify a deployment-strength reading of the reported numbers.

\textbf{Why these six.} P1--P3 control evidence-source confounds (window, sample, prior); P4--P6 control evidence-to-decision-mapping confounds (probability object, net cleansing, independence). The two groups partition the ``evidence $\to$ decision'' chain without overlap, and we restrict the list to failure modes that are LLM-specific or substantially amplified by LLMs (general-finance confounds such as factor drift are well-covered in the traditional quantitative literature).

\begin{table}[t]
  \caption{Six minimum-evidence protocols for LLM trading deployment claims. Each row is a single failure mode such that failing it disqualifies a deployment-strength reading of the reported numbers. Evidentiary anchors are listed inline in Sections 2--3.}
  \label{tab:protocols}
  \centering
  \footnotesize
  \renewcommand{\arraystretch}{1.25}
  \setlength{\tabcolsep}{5pt}
  \begin{tabular}{@{}>{\bfseries\raggedright\arraybackslash}p{0.115\linewidth} p{0.225\linewidth} p{0.345\linewidth} p{0.215\linewidth}@{}}
    \toprule
    \textbf{Protocol} & \textbf{Failure mode} & \textbf{Minimum reporting} & \textbf{If unmet} \\
    \midrule
    \multicolumn{4}{@{}l}{\textit{Group A. Evidence-source confounds}} \\
    \cmidrule(lr){1-4}
    P1. Temporal integrity
      & Time-travel alpha; pretraining or retrieval leakage; semantic future leakage
      & Model version, knowledge cutoff, post-training boundary, retrieval-corpus timestamps, memory-update rules; at least one post-cutoff or point-in-time window
      & At most historical-backtest evidence \\
    \addlinespace[2pt]\cmidrule(lr){1-4}\addlinespace[2pt]
    P2. Dynamic universe
      & Survivorship bias; ex-post-cleaned samples; static universes overstating returns
      & Time-varying tradable universe; delisting and suspension handling; liquidity filters; index-component changes; borrow and short-sale constraints
      & Alpha may come from ex-post-filtered universes \\
    \addlinespace[2pt]\cmidrule(lr){1-4}\addlinespace[2pt]
    P3. Counterfactual robustness
      & Parametric prior lock-in; insensitivity to reverse evidence; implicit sector or style tilts
      & Direction-flip rate, confidence shift, position-size shift under strong reverse evidence; sector-neutral and style-neutral prompt tests
      & Recommendations may reflect priors, not information \\
    \midrule
    \multicolumn{4}{@{}l}{\textit{Group B. Evidence-to-decision-mapping confounds}} \\
    \cmidrule(lr){1-4}
    P4. Epistemic calibration
      & Language confidence misread as trading probability; uncalibrated self-reported confidence
      & ECE, reliability curves, regime-conditioned calibration, out-of-sample calibration of any probability or confidence score used in sizing or risk control
      & LLM confidence should not control sizing \\
    \addlinespace[2pt]\cmidrule(lr){1-4}\addlinespace[2pt]
    P5. Realistic implementation
      & Gross-alpha illusion; accurate prediction with negative net return; token and latency costs cannibalizing returns
      & Layered gross-to-net cleansing: spread, slippage, commission, market impact, borrow cost, execution delay, token cost, inference latency
      & Profits cannot show deployability \\
    \addlinespace[2pt]\cmidrule(lr){1-4}\addlinespace[2pt]
    P6. Multi-agent disaggregation
      & Multi-agent consensus illusion; correlated errors across same-source models; echo-chamber and groupthink dynamics
      & Single-agent baseline, role similarity, disagreement rate, debate-round cost, coordination latency, multi-agent net-return delta
      & Debate is not independent-expert aggregation \\
    \bottomrule
  \end{tabular}
\end{table}

\textbf{Tiered applicability.} P1--P6 should apply in a tiered way according to claim strength:

\begin{table}[t]
  \caption{Tiered protocol requirements by claim strength. Stronger claims demand strictly stronger evidence: each row inherits the protocols of the row above, and the right column shows the language that remains defensible under that evidence floor.}
  \label{tab:tiered}
  \centering
  \footnotesize
  \renewcommand{\arraystretch}{1.30}
  \setlength{\tabcolsep}{6pt}
  \begin{tabular}{@{}>{\bfseries}p{0.21\linewidth} p{0.20\linewidth} p{0.51\linewidth}@{}}
    \toprule
    \textbf{Claim strength} & \textbf{Required protocols} & \textbf{Permissible language} \\
    \midrule
    LLM-as-text-extractor or research aid
      & P1 (light), P3 (light)
      & ``improves information extraction''; ``semantic features have marginal contribution'' \\
    \addlinespace[3pt]
    Historical backtest or prototype
      & P1 + P2 + P5 (minimum)
      & ``produces a positive-return trajectory in this window''; no deployment language \\
    \addlinespace[3pt]
    Deployable alpha claim
      & Full P1--P5
      & ``retains net return under structural tests'' \\
    \addlinespace[3pt]
    Autonomous trading ability
      & Full P1--P6
      & ``retains net return after multi-agent disaggregation'' \\
    \bottomrule
  \end{tabular}
\end{table}

\textbf{Reference implementation.} A reproducible audit harness covering P1 (post-cutoff window), P2 (dynamic universe), and P5 (full friction cleansing)---a one-year, five-ticker reproduction of TradingAgents and QuantAgent under the unified execution model and friction taxonomy of Figure~\ref{fig:friction}---is released at \url{https://github.com/hj1650782738/Trading}. Per-ticker gross/net metrics appear in Appendix B.

\section{A Modular Alternative: LLMs as Auditable Information Interfaces}
\label{sec:modular}

We are not proposing a new framework. This section describes the conservative modular practice that naturally results when the P1--P6 protocols of Section 4 are engineered into compliance: LLMs enter as auditable information interfaces upstream of independent calibration, risk, and execution modules, rather than as final trading-decision authority. This is consistent with~\citet{lopezlira2023chatgpt}, who show LLMs extract semantic signals correlated with short-term market reactions.

Standard quantitative pipelines have long separated alpha research, risk modeling, portfolio construction, and execution into independent modules with explicit interfaces and audits---this division predates LLMs and appears across active-portfolio textbooks and openly documented financial-ML platforms such as FinRL-Meta~\citep{liu2021finrlmeta} (which explicitly separates data, environment, and agent layers and encodes backtest rules and risk control as independent components). The pipeline runs in six stages---(1) information extraction, (2) feature construction, (3) signal synthesis, (4) probability calibration, (5) sizing and risk control, and (6) execution and audit---with a defensible LLM role concentrated at Stage~1 (schema-bound extraction from news, filings, calls) and tapering off into observer/explainer roles at Stages~5--6; full owner / role / audit-object decomposition is in Appendix E (Table~\ref{tab:modular}), which also walks an 8-K guidance cut through Stages~1--6 as a concrete example.

\textbf{Position implication.} The defensible role for LLMs in trading is as an auditable upstream interface, not as final decision authority. Authors should mark this boundary across the six stages before making deployment claims; reviewers should check that final authority does not rest on an uncalibrated LLM; deployment teams should ensure Stages 4--6 are owned by independent modules that can override LLM outputs.

\section{Alternative Views, Objections, and Limitations}
\label{sec:objections}

\textbf{We argue: reported alpha from end-to-end LLM trading systems should not be treated as evidence of deployable trading capability until it survives structural validity tests.} Sections 2--3 addressed technical objections inline; here we address paradigm-level objections that double as live limitations of our position---each concession marks a subclaim of ours that would be narrowed if the proponent supplied evidence under the relevant P1--P6 protocols.

\textbf{Concern 1.} We attack a strawman: most papers only claim backtest performance, not deployment readiness.
\textit{Response.} \emph{Partially correct.} We do not contest local ``research prototype'' framings. Our target is the slippage from reported returns into capability language---abstracts, talks, and downstream surveys repackaging the same numbers as ``LLMs can trade.'' P1--P6 are calibrated against deployment-grade claims, not against research existence.

\textbf{Concern 2.} Larger models, longer context, and stronger RAG will close the key information gaps.
\textit{Response.} \emph{Partially correct.} These improvements act primarily on the input layer; they do not automatically produce calibratable return probabilities, auditable position controls, or remove industry/size/style priors in model weights. \citet{merchant2025lookahead} indicate that RAG and prompt restrictions alone do not eliminate pretraining semantic memory. If, over a pre-specified post-cutoff horizon across multiple regimes, scale + RAG simultaneously improves ECE and net returns (P1, P4), our Section 3.1 subclaim should be narrowed.

\textbf{Concern 3.} Ensembles and multi-agent debate can cancel single-model bias.
\textit{Response.} \emph{Partially correct} only if error sources are sufficiently independent. \citet{zhang2025multiagent} report sub-$20\%$ debate win rates across $36$ configurations and \citet{henning2025notreplicate} show headline multi-agent gains do not consistently replicate across model families and seeds; the AMA live-trading benchmark~\citep{qian2025ama} further finds \emph{agent architecture}, not backbone choice, dominates outcome variation under identical execution rules---``add another LLM agent'' is not the same as ``add an independent expert.'' Authors claiming multi-agent improvement must report single-agent ablation, role similarity, disagreement rate, and net-return delta after controlling homogeneity and coordination cost (P5, P6).

\textbf{Concern 4.} LLMs have genuine financial-semantic value, so end-to-end approaches need not fail.
\textit{Response.} \emph{Orthogonal}---semantic value~\citep{lopezlira2023chatgpt} supports the modular practice in Section 5, not direct end-to-end deployment. If LLM-derived sentiment is genuinely predictive, it should enter downstream models as a feature with ablation, calibration, and post-friction validation.

\textbf{Concern 5.} Tool-calling and RL fine-tuning have already changed the architecture.
\textit{Response.} \emph{Partially correct.} FLAG-Trader~\citep{xiong2025flagtrader} fine-tunes a $135$M LLM with PPO and a Sharpe-increment reward and reports outperforming buy-and-hold on a small ticker set. We treat this as engineering progress, not refutation: the published evaluation does not yet jointly satisfy P1 (post-cutoff disclosure), P5 (full friction cleansing), and P3 (counterfactual robustness). The decisive question is whether final decision authority remains with an LLM that is not independently calibrated; if a future revision passes P3--P5, our Section 3 subclaim should be narrowed.

\textbf{Concern 6.} P1--P6 are too strict for early-stage prototypes.
\textit{Response.} \emph{Correct}---Table~\ref{tab:tiered} tiers requirements by claim strength. Prototypes need only avoid deployment language; an early-stage paper may report a positive-return trajectory provided it does not get promoted to ``deployable alpha.''

\textbf{Layered falsifiability.} Subclaim-level revisions: (i) significant post-friction net returns over a pre-specified post-cutoff horizon across regimes narrow Section 2; (ii) calibrated LLM confidence (pre-specified ECE) used for sizing narrows Section 3.1; (iii) significant multi-agent net-return deltas after controlling homogeneity and coordination cost narrow Concern 3.

\section{Conclusion}
\label{sec:conclusion}

The two-year arc of end-to-end LLM trading agents has been productive on its own terms: it forced concrete questions about how language models behave when asked to produce orderable decisions, and the resulting catalogue of architectures has real research value. The trouble starts when this exploratory progress is reported as deployment progress---headline Sharpe numbers from short, in-cutoff windows carried into abstracts, talks, and follow-up surveys as if they had already answered the deployment question.

Our argument has been narrow but specific. Public evidence cannot yet distinguish robust predictive ability from temporal contamination, unmodeled friction, short-sample noise, narrative fitting, and parametric prior; and even were every evaluation step cleaned up, three structural gaps would remain---language confidence is not tradable probability, narrative ability is not numerical execution, and model priors can become undisclosed implicit factor exposures. P1--P6 and the modular alternative are not a new framework but the minimum bookkeeping that lets a reader tell which kind of evidence a paper is offering---historical backtest, prototype, deployable claim, or autonomous trading ability---and lets reviewers hold each level of claim to its matching evidence floor.

We do not propose to police the academic literature; the exploratory contribution it has made is real and worth defending. We propose only to let it speak with the right scope. A backtest is a backtest, a prototype is a prototype, and a deployment claim is a deployment claim. Until reported alpha clears the structural tests of Section 4, it is more honestly read as exploratory evidence than as deployable trading capability---and the modular practice of Section 5 is the route that lets LLMs continue to enter trading systems without standing as the final source of trading authority.

\bibliographystyle{plainnat}
\bibliography{references}

\appendix

\section{Per-Ticker Reported Sharpe across Three LLM Trading Systems}
\label{app:per-ticker}

\begin{figure}[h]
  \centering
  \includegraphics[width=0.85\linewidth]{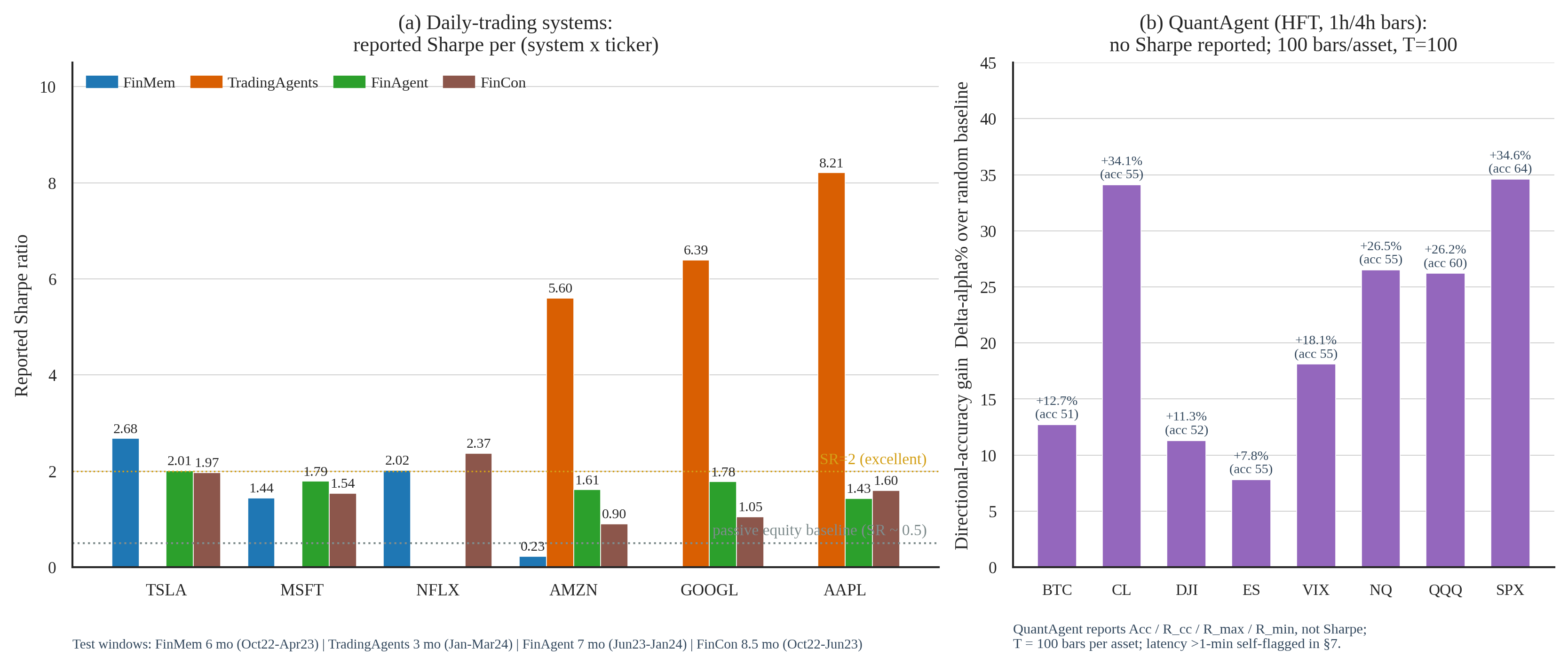}
  \caption{Reported Sharpe per (system $\times$ ticker) for four LLM trading systems. Bars show headline Sharpe as published by FinMem (6 mo, Oct 2022 -- Apr 2023), TradingAgents (3 mo, Jan -- Mar 2024), FinAgent (7 mo, Jun 2023 -- Jan 2024), and FinCon~\citep{yu2024fincon} (8.5 mo, Oct 2022 -- Jun 2023). All numbers are gross of friction except FinAgent, which models commission only.}
  \label{fig:system-ticker}
\end{figure}

For reference, Figure~\ref{fig:system-ticker} shows ticker-level headline Sharpe ratios as published by four LLM trading systems with comparable Sharpe-based reporting. QuantAgent~\citep{xiong2025quantagent} is omitted because it reports directional accuracy and short-window relative excess return on HFT instruments rather than daily-frequency Sharpe.

\section{Friction Coverage and Per-Asset Metrics from the Reproduction}
\label{app:portfolio}

\begin{figure}[h]
  \centering
  \includegraphics[width=0.95\linewidth]{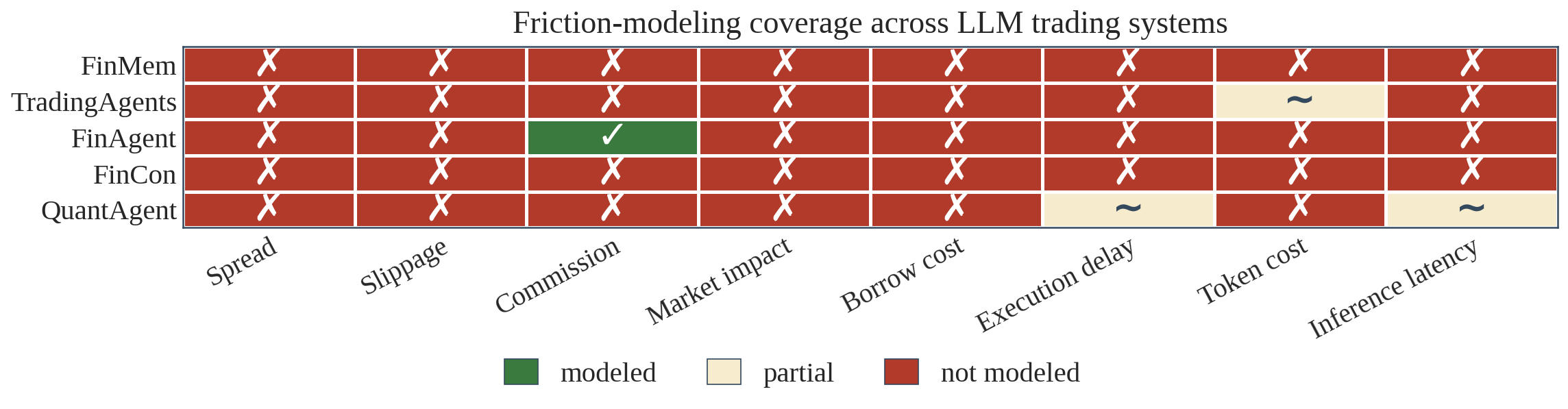}
  \caption{Friction-modeling coverage of FinMem, TradingAgents, FinAgent, FinCon, and QuantAgent across eight standard cost components.}
  \label{fig:friction}
\end{figure}

\paragraph{Friction coverage across systems.} Figure~\ref{fig:friction} maps the eight standard cost components onto five representative end-to-end LLM trading frameworks. Only commissions are explicitly modeled by all five; market impact, latency, slippage, financing, taxes, and token cost are absent in the majority of systems---$35$ of $40$ system\,$\times$\,friction cells unmodeled.

\paragraph{Per-asset breakdown of the empirical reproduction.} Table~\ref{tab:gross_net_results} reports the per-asset metrics underlying the portfolio curves of Figure~\ref{fig:portfolio} (Section 2.3). The setting is the same: \textbf{1-year backtest, 2025-01 to 2026-01, \$100K initial capital, equal-weight five-ticker portfolio (TSLA/NVDA/KO/XOM/MSTR); net charges commission, token cost, spread, and market impact}. Aggregated, buy-and-hold ends at $\$104.8$K; TradingAgents at $\$106.4$K gross / $\$102.3$K net; QuantAgent at $\$81.4$K gross / $\$77.9$K net.

\begin{table}[h]
\centering
\caption{Per-asset gross and net daily-trading metrics from our reproduction harness. CR: cumulative return; SR: Sharpe ratio; MDD: maximum drawdown; subscripts $g$/$n$: gross/net. The net column charges commission, token cost, spread, and market impact. The TCost column reports the commission component only (token cost is amortized at the portfolio level). Buy\&Hold net subtracts a one-time entry cost; TCost is omitted for display.}
\label{tab:gross_net_results}
\footnotesize
\setlength{\tabcolsep}{4pt}
\begin{tabular}{l|l|ccc|ccc|c}
\toprule
Asset & Agent & CR$_g$ (\%) & SR$_g$ & MDD$_g$ (\%) & CR$_n$ (\%) & SR$_n$ & MDD$_n$ (\%) & TCost (\$) \\
\midrule
 & Buy\&Hold & +9.61 & 0.45 & 49.41 & +9.56 & 0.45 & 49.44 & - \\
TSLA & TradingAgents & -2.01 & 0.23 & 55.50 & -10.17 & 0.09 & 58.33 & 1{,}632 \\
 & QuantAgent & -26.07 & -0.32 & 49.68 & -29.12 & -0.38 & 51.42 & 609 \\
\midrule
 & Buy\&Hold & +37.16 & 0.85 & 41.62 & +37.11 & 0.85 & 41.64 & - \\
NVDA & TradingAgents & +16.32 & 0.59 & 29.66 & +11.49 & 0.48 & 30.34 & 967 \\
 & QuantAgent & -56.07 & -1.79 & 56.07 & -59.30 & -1.91 & 59.28 & 646 \\
\midrule
 & Buy\&Hold & +12.82 & 0.76 & 11.24 & +12.77 & 0.76 & 11.25 & - \\
KO & TradingAgents & -10.84 & -0.87 & 17.89 & -17.17 & -1.42 & 22.34 & 1{,}265 \\
 & QuantAgent & -2.64 & -0.08 & 13.94 & -7.04 & -0.36 & 16.80 & 878 \\
\midrule
 & Buy\&Hold & +12.75 & 0.58 & 16.42 & +12.70 & 0.58 & 16.43 & - \\
XOM & TradingAgents & +11.26 & 0.55 & 16.39 & +10.71 & 0.54 & 16.40 & 111 \\
 & QuantAgent & -28.29 & -1.49 & 36.87 & -31.57 & -1.67 & 39.58 & 657 \\
\midrule
 & Buy\&Hold & -48.26 & -0.59 & 65.69 & -48.31 & -0.59 & 65.71 & - \\
MSTR & TradingAgents & +17.41 & 0.62 & 52.16 & +16.89 & 0.61 & 52.48 & 105 \\
 & QuantAgent & +20.15 & 0.61 & 44.96 & +16.62 & 0.56 & 46.35 & 707 \\
\bottomrule
\end{tabular}
\end{table}

\paragraph{What devours the gap between gross and net.} The gross-to-net distance for the two agent curves is the cumulative sum of the terms in equation~\eqref{eq:pnl_decomp}; the subset we charge---commission, token cost, spread, and market impact---turns CR$_g=-2.01\%$ into CR$_n=-10.17\%$ for TradingAgents on TSLA, and CR$_g=-10.84\%$ into CR$_n=-17.17\%$ on KO. Net buy-and-hold still beats both agents on $4$ of $5$ tickers; the only exception is MSTR, where buy-and-hold itself drew down $-48.31\%$. These numbers corroborate the central judgment of Section 2.3: \textbf{reported gross Sharpe is an upper bound, not a deployable number}. The same conclusion is reinforced by the Sharpe-uncertainty argument of Section 2.4---over one year ($\approx 252$ daily observations), the $95\%$ CI half-width of an agent's net Sharpe is large enough to span the entire buy-and-hold-vs-agent gap.

\section{Counterexamples and Classification of Representative Systems}
\label{app:counterexamples}

We classify systems / papers by their relation to our thesis: (a) apparent conflict but explainable via P1--P6; (b) compatible with our thesis; (d) approaching the modular solution and supportive. Among public evidence we have not yet found a definitive (c) challenge counterexample; if discovered, it would trigger the layered falsifiability conditions.

\begin{table}[h]
  \caption{Representative systems and their relation to the thesis.}
  \label{tab:representative}
  \centering
  \footnotesize
  \begin{tabular}{p{0.30\linewidth} c p{0.55\linewidth}}
    \toprule
    \textbf{System or paper} & \textbf{Class} & \textbf{Note} \\
    \midrule
    FinMem~\citep{yu2023finmem} and TradingAgents~\citep{xiao2024tradingagents} & (a) & Report short-window positive returns but lack P1 (cutoff), P5 (cleansing), and P6 (disaggregation); fits Section 2.4 \\
    FinCon~\citep{yu2024fincon} (NeurIPS 2024 main) & (a) & manager--analyst multi-agent + conceptual verbal reinforcement; per-ticker Sharpe $0.335$--$2.370$, portfolio Sharpe up to $3.269$; OOS window Oct 2022 -- Jun 2023 still constrained by P1 cutoff and P5 cleansing \\
    FinAgent~\citep{zhang2024finagent} & (a) & Multimodal inputs and tool calls but only deducts commission inside reward; remaining frictions unmodeled---see Figure~\ref{fig:friction} \\
    QuantAgent~\citep{xiong2025quantagent} & (a) and (d) & Reports headline Sharpe $1.76$--$2.02$ on HFT instruments---surface-level a counterexample to our thesis. On inspection, the architecture is mostly modular: the LLM emits structured reasoning over Indicator/Pattern/Trend/Risk roles, while execution and the $5$~bp stop-loss are off-LLM components. We therefore read it as already approximating the modular boundary advocated in Section 5, with residual P1 (cutoff disclosure) and P5 (full-friction net) gaps as the remaining deployment-evidence gaps. \\
    FLAG-Trader~\citep{xiong2025flagtrader} & (a) & RL fine-tune of a $135\text{M}$ LLM with PPO actor--critic and Sharpe-increment reward; reports SR $\approx 1.36$--$1.73$ on a small ticker set + BTC, but evaluation lacks explicit cutoff disclosure (P1) and friction cleansing beyond commission prompt mention (P5) \\
    FinBen~\citep{xie2024finben} (NeurIPS 2024 D\&B) & (b) & Holistic benchmark over 36 datasets and 24 tasks; GPT-4 FinTrade Sharpe $1.51 \pm 1.08$ (std exceeds half the mean) is a direct anchor for P1 + P5 \\
    FinRL-Meta~\citep{liu2021finrlmeta} & (d) & Already adopts modular division of labor, consistent with the inductive practice of Section 5 \\
    \citet{merchant2025lookahead} (Divergence Decoding) & (b) & Mechanistic anchor for P1: shows look-ahead bias is embedded in model parameters, supporting our argument rather than serving as a counterexample \\
    \bottomrule
  \end{tabular}
\end{table}

\section{Parametric Prior Lock-in: Detailed Same-Side Evidence}
\label{app:ppl-evidence}

This appendix expands the same-side evidence behind the monotonicity claim of Section 3. \citet{lee2025knowledge}'s counterfactual interventions show three patterns relevant to PPL:
\begin{itemize}
  \item Even when explicit reverse evidence is presented, models cling to existing sector tilts and style narratives---view-flip rates remain low and self-reported confidence does not contract proportionally.
  \item Models from the same ecosystem, pretraining corpus, or alignment paradigm exhibit highly converged sector and theme preferences, suggesting the prior is shared across what would otherwise be treated as ``independent'' agents.
  \item Persona prompts, multi-model voting, and multi-agent debate produce surface-level rhetorical differences but do not remove the underlying homogeneous prior; multi-agent agreement is therefore a poor proxy for independent expert agreement.
\end{itemize}

\begin{figure}[h]
  \centering
  \includegraphics[width=0.95\linewidth]{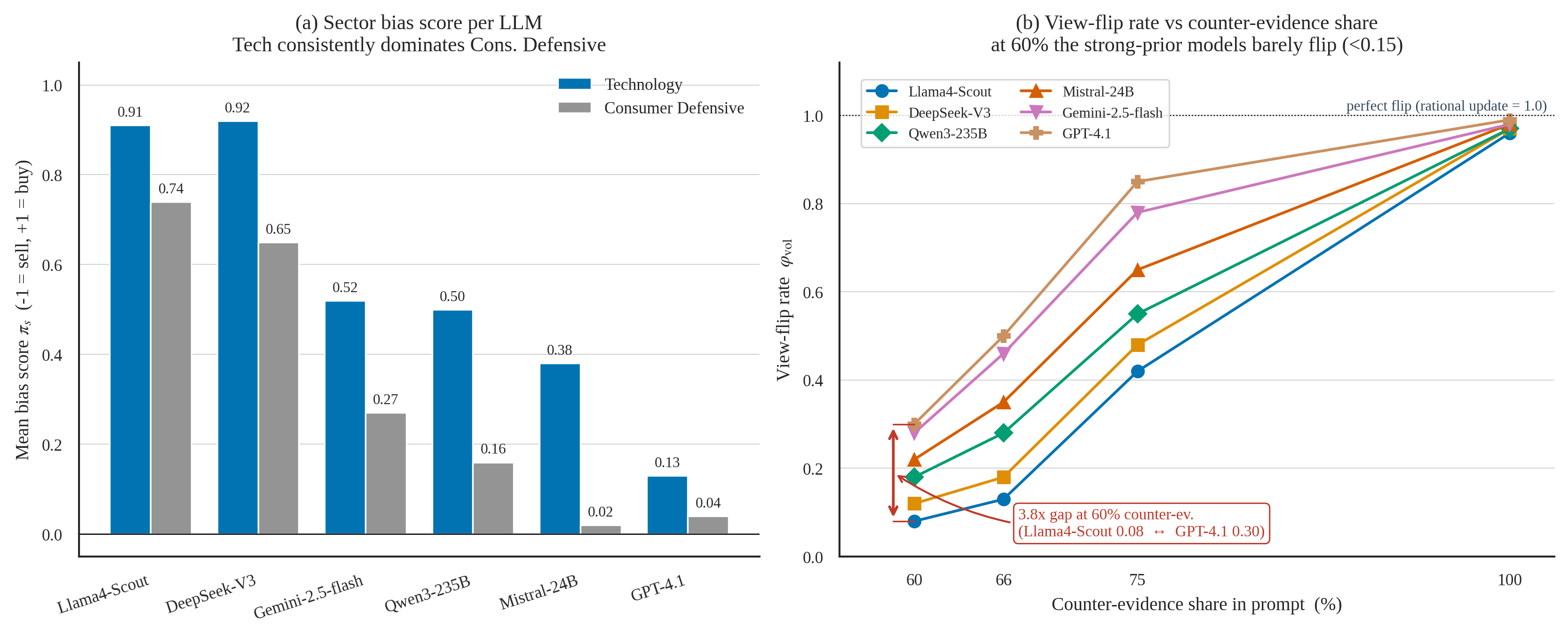}
  \caption{Two faces of parametric prior lock-in, after \citet{lee2025knowledge}. \textbf{(a)} Sector bias score $\pi_s \in [-1,+1]$ for the most- vs.\ least-favored sector (Technology vs.\ Consumer Defensive) across six LLMs; all six show a positive Tech tilt with a within-model gap that is statistically significant ($p<0.001$). \textbf{(b)} View-flip rate $\varphi_{\text{vol}}$ as a function of counter-evidence share, read off Lee et al.'s reported volume ratios $\{0|2,\,1|3,\,2|4,\,1|2,\,2|3\}$ corresponding to $\{100\%,\,75\%,\,66\%,\,66\%,\,60\%\}$ counter-evidence (the two $66\%$ ratios are averaged for clarity). At $100\%$ counter-evidence, all models flip near $1.0$; at $60\%$ counter-evidence the strong-prior models (Llama4-Scout, DeepSeek-V3) barely flip ($<0.15$), and the same models with the strongest sector tilts in panel~(a) sit lowest in panel~(b). The cross-axis correspondence is the empirical signature of PPL: a higher latent prior translates into a stickier posterior under reverse evidence.}
  \label{fig:ppl}
\end{figure}

\textbf{Quantitative anchors from~\citet{lee2025knowledge}.} The cross-model Tech-vs-Defensive bias-score gap (Figure~\ref{fig:ppl}a) ranges from $0.09$ (GPT-4.1: $0.13$ vs.\ $0.04$) to $0.36$ (Mistral-24B: $0.38$ vs.\ $0.02$), and is statistically significant for every model evaluated ($p<0.001$ on independent-samples $t$-test in Lee et al.\ Table~1). On the volumetric flip-rate side (Figure~\ref{fig:ppl}b), at the most adversarial ratio ($2|3$, i.e.\ $60\%$ counter-evidence), the strongest-prior model (Llama4-Scout) reverses its view in only $\approx 8\%$ of cases while GPT-4.1---the lowest-bias model---reverses in $\approx 30\%$, a $3.8\times$ gap that grows monotonically with the model's own sector-bias score in panel~(a). The same monotone correspondence appears under intensity-driven flips (Lee et al.\ Figure~7): even at $\Delta = 10$ (counter-evidence claimed price impact $2\times$ supporting), the high-bias models stay below $\varphi_{\text{int}} = 0.6$, well below ``perfect rational update.'' Panels~(a) and~(b) jointly motivate treating direction-flip rate, confidence shift, and position-size shift as the three monotonicity axes of P3.

These patterns motivate treating P3 (counterfactual robustness) as a core protocol rather than an optional ablation: a system whose recommendations do not move under reverse-evidence prompts cannot be distinguished from one operating on parametric prior alone.

\section{Worked Example: 8-K Guidance Cut Through the Modular Pipeline}
\label{app:worked-example}

\begin{table}[h]
  \caption{Common stage structure of modular quantitative trading systems (referenced from Section 5).}
  \label{tab:modular}
  \centering
  \footnotesize
  \begin{tabular}{p{0.16\linewidth} p{0.20\linewidth} p{0.28\linewidth} p{0.28\linewidth}}
    \toprule
    \textbf{Stage} & \textbf{Owner} & \textbf{Reasonable LLM role} & \textbf{Audit object} \\
    \midrule
    1. Information extraction & LLM-led, schema-bound & Extract structured propositions from news, filings, calls, announcements & Source span, publication time, retrieval time, entity/event labels \\
    2. Feature construction & Quant module & Provides candidate structured inputs, does not set feature weights & Point-in-time feature table, missing-data handling, normalization rules \\
    3. Signal synthesis & Quant model & One of many information sources & Ablation and marginal contribution of LLM-derived features \\
    4. Probability calibration & Independent statistical module & Does not provide self-rated trading probabilities & ECE, reliability curves, regime-conditioned calibration, out-of-sample stability \\
    5. Sizing and risk control & Portfolio and risk modules & Explains sources of risk, does not directly determine size & Factor exposures, sector exposures, leverage, drawdown, liquidity constraints \\
    6. Execution and audit & Execution system & Observer or summarizer & Order log, fill prices, slippage, latency, failed orders, risk-stop records \\
    \bottomrule
  \end{tabular}
\end{table}

This appendix walks the modular stage structure (Table~\ref{tab:modular}) on a concrete event, as a description of existing practice rather than a proposal. Suppose at time $t$ an 8-K is released stating that company $X$ is cutting next-quarter revenue guidance.

\begin{itemize}
  \item \textbf{Stage 1} (LLM extraction): emits JSON \texttt{\{entity: X, event: 'guidance\_cut', evidence\_quote: '...', source\_time: t, retrieval\_time: t+$\delta$\}}.
  \item \textbf{Stage 2} (feature construction): the quant system maps it to a surprise feature, joined with historical EPS surprise and guidance-revision history.
  \item \textbf{Stage 3} (signal synthesis): the quant model measures IC and marginal contribution on multi-year panel data.
  \item \textbf{Stage 4} (calibration): an independent calibration module assesses out-of-sample ECE.
  \item \textbf{Stage 5} (sizing and risk control): the portfolio module enforces sector-neutral risk-budget caps.
  \item \textbf{Stage 6} (execution and audit): the execution system records timestamps, slippage, and risk-control intercepts.
\end{itemize}

The LLM leads Stage 1; in subsequent stages it is at most an observer or explainer. The point of the example is not the architecture itself---which is standard---but that compliance with P1--P6 (cutoff disclosure, calibration, friction cleansing, multi-agent disaggregation) is achieved \emph{by construction} once final decision authority is moved off the LLM.


\end{document}